\documentclass[aps,twocolumn,prd,showpacs,floatfix]{revtex4}

\usepackage{mathrsfs}
\usepackage{dcolumn}
\usepackage{bm}
\usepackage{amsmath}
\usepackage{amsthm}
\usepackage{amssymb}
\usepackage{graphicx}
\usepackage{footmisc}
\usepackage{nicefrac}
\usepackage{appendix}

\newcommand{\eref}[1]{(\ref{#1})}
\newcommand{\eps }{\varepsilon }

\begin{document}

\title{Narrow resonances and black-hole-like absorption in a non-black-hole
metric}
\author{V. V. Flambaum}
\affiliation{School of Physics, University of New South Wales, Sydney 2052,
Australia}
\author{G. H. Gossel}
\affiliation{School of Physics, University of New South Wales, Sydney 2052,
Australia}
\author{G. F. Gribakin}
\affiliation{Department of Applied Mathematics and Theoretical Physics,
Queen's University, Belfast BT7 1NN, Northern Ireland, UK}

\date{\today}

\begin{abstract}
A massive body with the Schwarzschild interior metric (perfect fluid of
constant density) develops a pressure singularity at the origin when the radius
of the body $R$ approaches $9r_s/8$, where $r_s$ is the
Schwarzschild radius. We show that a quantum scalar particle scattered in this
gravitational field possesses a dense spectrum of narrow resonances. 
Their density and lifetimes tend to infinity in the
limit $R\rightarrow 9r_s/8$, and we determine the cross section of the particle
capture into these quasibound states. Therefore, a body that is not a black hole
demonstrates black-hole-like absorption.
\end{abstract}


\maketitle  

\section{Introduction}
We found in our previous work \cite{Resonances} that the scattering of
a massless scalar particle in a near-black-hole metric of a massive body
whose radius $R$ slightly exceeds the Schwarzschild radius $r_s$, is
characterized by a dense spectrum of resonances. For $R \to r_s$ both
the resonance energy spacing $D$ and their width $\gamma$ tend to zero, while
their ratio remains finite ($\gamma/D \simeq 2 \eps^{2}r_s^2/\pi$) and tends
to zero for small energies $\eps $.
(We use units where $\hbar =c =1$.) This allowed us to define the cross section
for particle capture into these long-lived states in the spirit of the optical
model \cite{LLV3}, by averaging over a small energy interval containing many
resonances. Note that this capture emerges in a purely potential scattering
problem, without any absorption introduced {\em a priori}. Somewhat
unexpectedly, the capture cross section turned out to be equal to the
cross section obtained by assuming total absorption at the event horizon.

In particular, in the zero-energy limit our result coincides with Unruh's
absorption cross section $\sigma _a=4\pi r_s^2$ for a black hole
\cite{Unruh,Matzner, Starob, Sanchez,Das,Crispino,Decanini}.
This shows that a non-singular static metric can acquire black-hole
properties prior to the actual formation of the black hole.

It is interesting to see whether this resonance absorption phenomenon is
specific to near-black-hole metrics, or if there are other instances of 
similar behaviour. For example, consider a massive star modelled as an
incompressible fluid sphere. The interior of such a body is described by
the Schwarzschild interior metric \cite{SchwarzsInt}. This metric develops a
pressure singularity at the origin for $R\rightarrow 9r_s/8$, far form the
black-hole limit. In Ref.~\cite{GosseBoundStates} we examined the spectrum of
discrete bound states for a massive scalar particle in the gravitational
field described by this metric. It was found that near the singularity this
spectrum acquires some black-hole-like
features. In particular, all the levels with finite principal quantum number
$n$ collapse towards zero energy (binding energy is $-mc^2$), i.e., the
spectral density becomes infinite, similar to that of a near-black-hole
metric (cf. Ref. \cite{Soffel}). 
 
This gives us a motivation to search for narrow resonances in the Schwarzschild
interior metric. Indeed, in the present work  we find such resonances in the
limit  $R\rightarrow 9r_s/8$. However, in contrast to the black-hole-like case
considered in Ref. \cite{Resonances}, the effective potential produced by the
Schwarzschild interior metric possesses a barrier near the boundary, and
the long resonance lifetimes are due to the particle having to tunnel
through this barrier to the surface of the body.
 
Similar to the black-hole-like behaviour for $R\rightarrow r_s$,
the resonance width $\gamma$ and energy spacing $D$ tend to zero for the
Schwarzschild interior metric in the limit  $R\rightarrow 9r_s/8$, while their
ratio $\gamma/D$ remains finite and tends to zero at small energies $\eps $.
One can thus define the ``optical'' capture cross section by averaging over a
small energy interval containing many resonances. In contrast to the
near-black-hole case, where the cross section remains finite, the
absorption cross section for the Schwarzschild interior metric tends to zero
at zero energy. However, this metric still possesses black-hole-like
absorption for nonzero energies. This  means that such an object may
gravitationally absorb particles in a black-hole-like manner despite the fact
that its metric never approaches that of a black hole.

In what follows we analyse the massless scalar particle scattering problem for
the Schwarzschild interior metric both numerically and analytically.
Although the latter treatment of the problem is only approximate, it confirms
all the important features of the emerging resonant scattering picture.

\section{Radial Klein-Gordon equation}\label{sec:KG}

The Klein-Gordon equation for a scalar particle of mass $m$ in a curved
space-time with the metric $g_{\mu \nu}$ is
\begin{equation}\label{eq:KG}
\partial_\mu (\sqrt{-g} g^{\mu\nu}\partial_\nu \Psi)+\sqrt{-g} m^2\Psi =0.
\end{equation}
For a particle of energy $\eps $ in a spherically symmetric field we seek
solution of Eq.~(\ref{eq:KG}) in the form
$\Psi (x)=e^{-i\eps t}\psi (r)Y_{lm}(\theta ,\varphi )$. Considering
for simplicity the case of a massless particle with zero angular momentum
($l=0$) and the metric of the form
\begin{equation}\label{eq:ds2}
\mathrm{d}s^2 =a(r) dt^2-b(r)dr^2 -r^2 d\Omega^2,
\end{equation}
where $d\Omega ^2=d\theta ^2+\sin ^2\theta d\varphi^2 $, the radial wave
equation is given by
\begin{equation}\label{eq:GenWave1}
\psi''(r)+ \left[ \frac{2}{r}+\frac{h'(r)}{h(r)}\right]\psi'(r) +
\frac{\eps^2}{h^2(r)}\psi(r) = 0
\end{equation}
where $h(r) = \sqrt{a(r)/b(r)}$.
By transforming the radial function as
$\psi(r) =r^{-1}\phi(r)/\sqrt{h(r)}$, Eq.~(\ref{eq:GenWave1}) can be cast in
the following Schr\"odinger-like form,
\begin{equation}\label{eq:GenSchrod}
\phi''(r) +\left\{ \frac{\eps^2}{h^2}+\frac{1}{4}\left[\frac{h'}{h}\right]^2
-\frac{h''}{2 h}-\frac{h'}{r h}\right\} \phi(r) = 0.
\end{equation}
 
This equation is convenient for deriving its semiclassical (WKB) solution.
For small $h(r)$ (near $r=0$, see below) the first
term in braces dominates and this solution is
\begin{equation}\label{eq:SCSolGen}
\phi(r)=\sqrt{\frac{h(r)}{\eps}}\sin\left(\eps\int_{0}^{r}\frac{dr}{h(r)}
\right).
\end{equation}
In fact, it is easy to verify that this solution is \textit{exact} if the
$-h'/rh$ term in braces in Eq.~\eref{eq:GenSchrod} is neglected.


\section{Phaseshift for the Schwarzschild interior metric}
\label{sec:GeneralCase}

Outside a spherically symmetric, non-rotating body of mass $M$ and
radius $R$ the metric is given by the Schwarzschild solution
\begin{equation}\label{eq:ExteriorMetric}
ds^2 = \left(1-\frac{r_s}{r}\right)dt^2-\left(1-\frac{r_s}{r}\right)^{-1}dr^2 -
r^2 d\Omega^2 ,
\end{equation}
where $r_s = 2GM$ is the Schwarzschild radius of the body and $G$ is the
gravitational constant. Hence, for $r>R$ Eq.~\eref{eq:GenWave1} takes the form
\begin{equation}
\label{eq:ExteriorWave}
\psi''(r) + \left(\frac{1}{r-r_s}+\frac{1}{r}\right)\psi'(r)
+\frac{r^2 \eps ^2}{(r-r_s)^2}
\psi (r) = 0.
\end{equation}
The Schwarzschild {\em interior} solution describes a static spherical
mass of incompressible perfect fluid with constant density, and
is given by Eq.~(\ref{eq:ds2}) with
\begin{align*}\label{eq:ab}
a(r)&=\left(\frac{3}{2}\sqrt{1-\frac{r_s}{R}}-
\frac{1}{2}\sqrt{1-\frac{r_s r^2}{R^3}}\right)^2,\\
b(r)&=\left(1-\frac{r_s r^2}{R^3}\right)^{-1},
\end{align*}
so that
\begin{equation}\label{eq:SchwarzschildInterior}
h(r) =\frac{1}{2} \sqrt{1-\frac{r_sr^2}{R^3}}\left(3\sqrt{1-\frac{r_s}{R}}-
\sqrt{1-\frac{r_sr^2}{R^3}}\right).
\end{equation}
This metric is valid for $r_s<8R/9$ but develops a singularity as $a(0)$
vanishes for $r_s=8R/9$ \citep{Buchdahl}.

Using $h(r)$ from  Eq.~(\ref{eq:SchwarzschildInterior}) in
Eq.~(\ref{eq:GenWave1}) gives the radial wave equation for $r<R$. For
$r_s < 8R/9$ this metric is smooth at the origin. Hence, the regular solution
for $l=0$ must satisfy the boundary conditions $\psi'(0)=0$, $\psi(0)\neq 0$.
The value of $\psi(0)$ only affects the normalization of the wavefunction and
we set $\psi(0)=1$. We solve the interior equation numerically
using {\em Mathematica} \cite{math}.

This solution supplies the boundary condition for the {\em exterior}
wavefunction at the surface of the body $r=R$. ($R=1$ is used in the numerical
calculations.) Equation (\ref{eq:ExteriorWave}) is
then integrated outwards to large distances $r\gg r_s$. In this asymptotic
region~Eq. (\ref{eq:ExteriorWave}) takes the form of the nonrelativistic
Shr\"odinger equation for a particle with momentum $\eps $ and unit mass
in the attractive Coulomb potential with charge $Z=-r_s\eps ^2$. Here the
wavefunction can be matched with the asymptotic Coulomb solution \cite{LLV3},
\begin{equation}\label{eq:CoulombMatch}
\psi (r)\propto \sin [\eps r - (Z/\eps ) \ln 2 \eps r +\delta_C+
\delta ],
\end{equation}
where $\delta_C = \arg \Gamma (1+i Z/\eps ) $ is the Coulomb phaseshift,
to find the short-range phaseshift $\delta $. The latter is determined
almost exclusively by the interior metric, and carries important information
about the behaviour of the wavefunction at $r<R$.

\begin{figure}[t!]
\begin{center}
\includegraphics[width=0.48\textwidth]{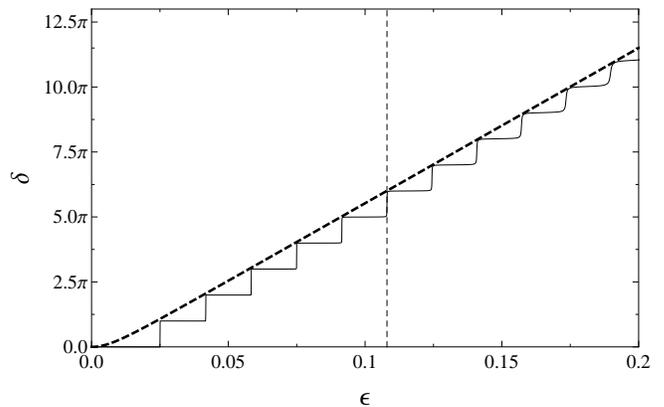}
\caption{The short-range phaseshift $\delta $ obtained numerically as a
function of $\epsilon =\eps R$ (solid line) and the semiclassical phase,
Eq.~(\ref{eq:SCPhase}), at the classical turning point (thick dashed line) for
$\xi = 8/9-r_s/R = 0.00014$. The vertical dashed line corresponds to 
the sixth resonance at $\epsilon \approx 0.1080$.
\label{fig:TotalPhase}}
\end{center}
\end{figure}

\begin{figure}[t!]
\begin{center} 
\includegraphics[width=0.48\textwidth]{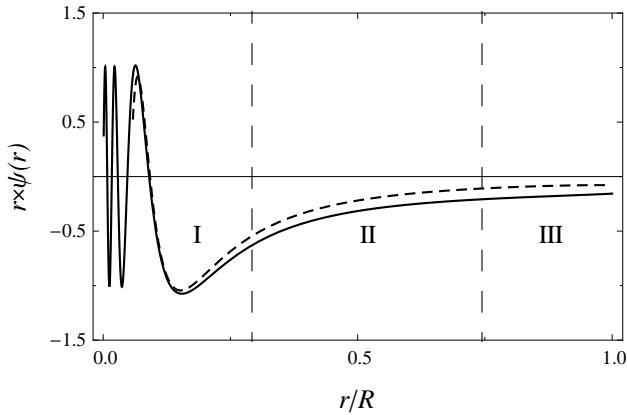}
\caption{The radial wavefunction at the energy $\epsilon =0.1080$ of the
sixth resonance for $\xi = 0.00014$ (solid curve). The dashed curve is the
analytical approximation for the wavefunction given by Eqn.~\eref{eq:ExactSol}
for the energy $\eps_n$ ($n=6$) from Eqn.~\eref{eq:Energy}. The vertical lines
indicate the two classical turning points defined by $p(r) = 0$. Regions I and
III are classically allowed, while II is classically forbidden.
\label{fig:Wavefunction}}
\end{center}
\end{figure}

Unlike $\delta _C$ which is small and has a weak dependence on the energy, the
phaseshift $\delta $ becomes large when $r_s$ approaches $8R/9$, i.e., for
\begin{equation}\label{eq:xi}
\xi \equiv 8/9-r_s/R \ll 1.
\end{equation}
This phaseshift also has a strong dependence on the energy of the particle, as
shown by the solid line in Fig.~\ref{fig:TotalPhase}. Similar to the case of a
near-black-hole metric considered in Ref.~\cite{Resonances}, this phaseshift
goes through many steps of the size $\pi$. Each of these steps corresponds to
a resonance, i.e., a long-lived quasibound states of the projectile and the
target. For energies corresponding to the midpoints of the steps (where the
derivative $d\delta /d\eps $ is largest) the magnitude of the wavefunction
$\psi (r)$ inside the body ($r<R$) is much greater than outside. This is a
signature of a quasibound state. The wavefunction at one of the resonances
is shown in Fig.~\ref{fig:Wavefunction}.

The resonances observed in the phaseshift $\delta $ are charterised by
their energies $\eps _n$ and widths $\gamma _n$, which can be determined
numerically by fitting the ``steps'' in $\delta $ by
$\arctan [2( \eps - \eps _n)/\gamma _n]+\mbox{const.}$ \cite{LLV3}, or
analytically (see Sec.~\ref{sec:an}). We show below
(see Figs. \ref{fig:EnergyPlot} and \ref{fig:WidthPlot}) that as the
Schwarzschild radius $r_s$ tends to $8R/9$, i.e., for
$\xi \rightarrow 0$, both the energies and the widths of the
resonances tend to zero (i.e., their lifetimes tend to infinity).
This means that in this limit the massive body with the Schwarzschild interior
metric develops absorption properties usually typical of black holes.


\section{Energies and widths of the resonances}\label{sec:an}

\subsection{Schwarzschild interior}
When the wave equation is written in the Schr\"odinger-like form
(\ref{eq:GenSchrod}), the coefficient in braces playes the role of
an effective potential for the motion a particle with the classical momentum
\begin{equation}\label{eq:p}
p(r)= \left\{\frac{\eps^2}{h^2}+\frac{1}{4}\left[\frac{h'}{h}\right]^2
-\frac{h''}{2 h}-\frac{h'}{r h}\right\}^{1/2}.
\end{equation}
In the case of the Schwarzschild interior metric, $h(r)$ is given by
Eq.~\eref{eq:SchwarzschildInterior}. For $\xi \ll 1$ and $r\ll R$ (in practice
$r<0.5R$ is sufficient) $h(r)$ can be approximated as follows,
\begin{equation}\label{eq:hap}
h(r)\simeq \frac{9\xi }{4}+\frac{2r^2}{9R^2}-\frac{4r^4}{81R^4}.
\end{equation}
This shows that for small $r$, where the first term in braces
in Eq.~(\ref{eq:p}) dominates, the particle has a large classical momentum
(e.g., $p(0)\simeq 4\eps /9\xi $). This corresponds to the motion
in a deep classically allowed region near the origin.
Analysis of Eqs.~(\ref{eq:SchwarzschildInterior}) and (\ref{eq:p}) shows that
for $\xi \ll 1$ there is a second classically allowed region near the
boundary with a broad, nonsemiclassical potential barrier in between,
cf. Fig.~\ref{fig:Wavefunction}. These regions are separated by
two classical turning points $r_1$ and $r_2$ given by the roots of $p(r)=0$.

The above picture explains the origins of the dense spectrum of narrow
resonances observed in the limit $\xi \rightarrow 0$. To estimate the resonance
widths analytically we require an expression for the scattering matrix, which
is constructed using the procedure outlined in the following subsections.

\subsubsection{Region I: origin to barrier}\label{subsubsec:I}

In this region  ($0\leq r \ll r_1$) the $\eps ^2/h^2$ term in
Eq.~\eref{eq:p} dominates near the origin for $\xi \ll 1$.
Dropping the last term in Eq.~(\ref{eq:hap}) we obtain the
semiclassical phase of the wavefunction $\phi(r)$, Eq.~(\ref{eq:SCSolGen}),
\begin{align}\label{eq:SCPhase}
\int_{0}^{r} \frac{\eps}{h(r)}dr &\simeq \eps \int_{0}^{r}
\left(\frac{9\xi }{4}+\frac{2r^2}{9R^2}\right)^{-1} dr \notag{} \\ 
&= \frac{\epsilon \sqrt{2}}{\sqrt{\xi }}
\arctan \left(\frac{\rho \sqrt{8}}{9\sqrt{\xi }}\right),
\end{align}
where $\epsilon =\eps R$ and $\rho =r/R$ are the scaled energy and radial
coordinate, respectively. The wavefunction in this region is then
\begin{equation} \label{eq:Sol1}
\phi_\mathrm{I}(\rho ) =C\sqrt{\frac{9\xi}{4}+\frac{2\rho ^2}{9}}
\sin \left[\frac{\epsilon \sqrt{2}}{\sqrt{\xi}}\arctan 
\left(\frac{\rho \sqrt{8}}{9\sqrt{\xi }}\right)\right],
\end{equation}
where $C$ depends on normalization.

\subsubsection{Region II: suppression by barrier}
In the intermediate range, $\sqrt{\xi }\ll \rho  \lesssim 0.5$,
the square of the classical momentum (\ref{eq:p}) simplifies to
\begin{equation} \label{eq:VeffII}
p^2(r)\simeq \frac{81R^4 \eps ^2}{4r^4}-\frac{2}{r^2}.
\end{equation}
In this case Eq.~(\ref{eq:GenSchrod}) has an exact analytical
solution,
\begin{equation}\label{eq:ExactSol}
\phi_{\mathrm{II}}(\rho ) = A \rho \sqrt{81\epsilon ^2 +4\rho ^2} 
\sin\left[\Phi+\frac{9\epsilon }{2\rho }+
\arctan\left(\frac{2\rho }{9\epsilon}\right)\right],
\end{equation}
where $\Phi$ and $A$ are constants determined by matching to the
solution (\ref{eq:Sol1}) to the left of the first classical turning point
$\rho _1=9\epsilon /\sqrt{8}$ [from Eq.~(\ref{eq:VeffII})].

For $\sqrt{\xi} \ll \rho <\rho _1$ both Eq.~\eref{eq:Sol1} and
Eq.~\eref{eq:ExactSol} are valid (the former is valid at the origin while the
latter is not). Matching is done by first expanding the $\arctan $ term
in Eq.~\eref{eq:Sol1} for $\rho /\sqrt{\xi}\gg 1$, and neglecting the
$\arctan $ term in Eq.~\eref{eq:ExactSol}, which is justified for
$2\rho /9\epsilon \ll 1$. This yields the phase
\begin{equation}\label{eq:Phi}
\Phi = -\frac{\pi \epsilon }{\sqrt{2\xi }}.
\end{equation}


When the energy is on resonance, the wavefunction decreases
under the barrier (Region II in Fig.~\ref{fig:Wavefunction}). Considering
Eq.~\eref{eq:ExactSol} for $\rho \gg \rho _1$ (but keeping
$\rho <0.5$, which is possible for low energies $\epsilon \ll 1$), and
expanding the $\arctan $ function for large arguments, we have
\begin{equation}\label{eq:IntNearBoundary}
\phi _{\mathrm{II}}(\rho )\simeq -2A \rho ^2 \sin\left[
\frac{\pi \epsilon }{\sqrt{2\xi }}-\frac{\pi}{2}-
\frac{1}{3}\left(\frac{9\epsilon }{2\rho }\right)^3\right].
\end{equation}
This function represents a decreasing solution
$\phi_{\mathrm{II}}(\rho) \propto \rho^{-1}$ only if we require
$\pi \epsilon/\sqrt{2\xi}-\pi/2 = n \pi$ ($n = 1, 2, \dots$), which gives the
resonance energies 
\begin{equation}\label{eq:Energy}
\epsilon _n = \sqrt{2\xi} (n+1/2).
\end{equation}
Hence the resonances form an equispaced spectrum similar to that of
a harmonic oscillator (starting from $n=1$ though). This is in agreement with
the behaviour of the phaseshift in Fig.~\ref{fig:TotalPhase}.

\begin{figure}[t!]
\begin{center}
\includegraphics[width=0.48\textwidth]{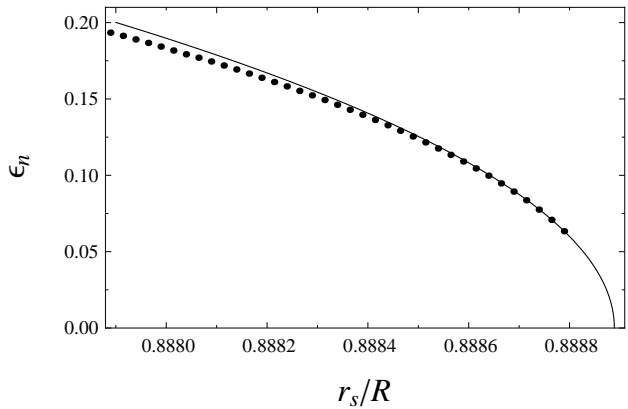}
\caption{Energies of the $n=4$ resonance given by Eq.\eref{eq:Energy} (solid
line) and obtained numerically by fitting the resonance jumps of the phaseshift
$\delta $ (solid circles) as a function of $r_s/R$.
\label{fig:EnergyPlot}}
\end{center}
\end{figure}

According to Eq.~(\ref{eq:Energy}), for $\xi \rightarrow 0$ the energies of
all resonances tend to zero. This behaviour is shown in
Fig.~\ref{fig:EnergyPlot} which compares the analytical expression
for $\epsilon _n$ with the values obtained numerically from the phaseshift
$\delta $.

\subsubsection{Region III: barrier to boundary}

To calculate the $S$-matrix and determine the resonance widths analytically
we need to match the interior and exterior solutions at the boundary $r=R$.
Formally, the effective potential in Eq.~\eref{eq:GenSchrod} contains a
second classical turning point $r_2$, which defines the outer classically
allowed region, $r_2<r<R$. In this region the classical momentum remains small
and the wavefunction changes little between $r_2$ and $R$ (see
Fig.~\ref{fig:Wavefunction}). Indeed, it is easy to check that the classical
momentum takes its largest values at the boundary $r=R$, and for
$\xi \rightarrow 0$ one has $p(R)=\sqrt{81\eps ^2+58R^{-2}}$. This momentum
is much smaller than the momenta at the origin (see Sec.~\ref{subsubsec:I}) or
at small $r$ [see Eq.~(\ref{eq:VeffII})].

Therefore in generating an approximate analytical solution over the entire
interior region we assume that the major contributions arise from regions I and
II. Changes in the wavefunction due to the potential in region III give rise
to higher-order effects which we calculate as corrections at the end of this
section. 
Thus we formally extend the solution in Eq.~\eref{eq:ExactSol} to the boundary
$\rho =1$ where it is matched to an appropriate solution of the exterior
equation as discussed below.

\subsection{Schwarzchild exterior solution}

For low-energy scattering, $\eps R\ll 1$, the last term in the exterior
equation (\ref{eq:ExteriorWave}) can be neglected near the boundary.
Hence, for $r\sim R$, the solution of Eq.~(\ref{eq:ExteriorWave}) is
\begin{equation}\label{eq:alb}
\psi (r) = \alpha \ln\left(\frac{r-r_s}{r}\right)+\beta ,
\end{equation}
where $\alpha $ and $\beta $ are constants determined by joining this solution
with the interior solution at $r=R$ (see below). Following the matching
procedure outlined in Ref.~\cite{ScalarScattering} the low-energy $s$-wave
scattering matrix is found as $S_{0}=\tilde S _0 e^{2i\delta_C}$, where
\begin{equation} \label{eq:SMatrix}
\tilde S_{0}\equiv e^{2i\delta } =\frac{(\beta /\alpha )-i \eps r_s C^2}
{(\beta /\alpha )+i \eps r_s C^2}
\end{equation}
is the short-range part of the scattering matrix, and 
$C^2=2\pi\eps r_s /[1-\exp(-2\pi \eps r_s)]$ is a function 
that characterizes the long-range Coulomb tail of the potential \cite{ScalarScattering}.

In general, resonances can be found as poles of the $S$-matrix in the
complex energy plane at $\eps =\eps _n-i\gamma _n/2$ \cite{LLV3}. On the real
energy axis they correspond to the points where the short-range phaseshift
$\delta $ passes through $\pi (n-1/2)$. At these energies $\tilde S_0=-1$,
which requires $\beta/\alpha = 0$. This means that near the resonances
one can use
\begin{equation}\label{eq:betal}
\beta /\alpha \simeq (\beta /\alpha )'(\eps-\eps_n ),
\end{equation}
where the prime denotes the derivative with respect to $\eps$ taken at the
resonance. Using this expression and looking for the poles of $\tilde S_0$,
Eq.~(\ref{eq:SMatrix}), gives the resonance width
\begin{equation}\label{eq:Width}
\gamma_n =2\frac{r_s \eps_n C^2}{(\beta/\alpha)'}.
\end{equation}

To find the ratio $\beta /\alpha $ and $(\beta/\alpha)'$ we match the
logarithmic derivatives of the interior and exterior wavefunctions at the
boundary. Using the relation between the solutions $\psi (r)$ and 
$\phi(r) $ (Sec.~\ref{sec:KG}) for the interior metric,
Eq.~(\ref{eq:SchwarzschildInterior}), the logarithmic derivative
(evaluated at the boundary) is 
\begin{align}\label{eq:IntLogDeriv}
\left.\frac{\psi '(r)}{\psi (r)}\right|_{R-}
&= \frac{1}{R} + \left.\frac{\phi'(r)}{\phi (r)}\right|_{R-} \notag{} \\
&=\frac{3}{R}-\frac{1}{R}\left(\frac{9\epsilon }{2}\right)^3 
\tan\left[\frac{\pi \epsilon }{\sqrt{2\xi}}-
\frac{1}{3}\left(\frac{9\epsilon }{2}\right)^3\right],
\end{align}
where we used Eq.~(\ref{eq:ExactSol}) to obtain the last line.

The exterior logarithmic derivative at the boundary for
$r_s \rightarrow 8/9R$ is found from Eq.~(\ref{eq:alb}) as
\begin{equation}
\label{eq:ExtLogDeriv}
{\frac{\psi '(r)}{\psi (r)}}\bigg|_{R+} = \frac{8\alpha}{R (\beta-\alpha \ln9)}.
\end{equation}
Setting the logarithmic derivatives from Eq.~\eref{eq:IntLogDeriv} and
Eq.~\eref{eq:ExtLogDeriv} equal, gives 
\begin{equation}
\frac{\beta}{\alpha}=\frac{8}{R\left[ \psi'(r)/\psi (r)\right]_{R-}}+ \ln 9.
\end{equation}
When finding $(\beta /\alpha )'$ to estimate $\gamma_n$ from
Eqn.~\eref{eq:Width}, we can use the fact that at (or very near to) the
resonance, $\beta /\alpha = 0$.  This allows one to simplify the
answer and obtain
\begin{equation} \label{eq:Widths}
\gamma_n = B \sqrt{\xi}\,\eps_{n}^4 R^3,
\end{equation}
with the constant $B$ given approximately by $B=2.74$.

\begin{figure}[t!]
\begin{center}
\includegraphics[width=0.48\textwidth]{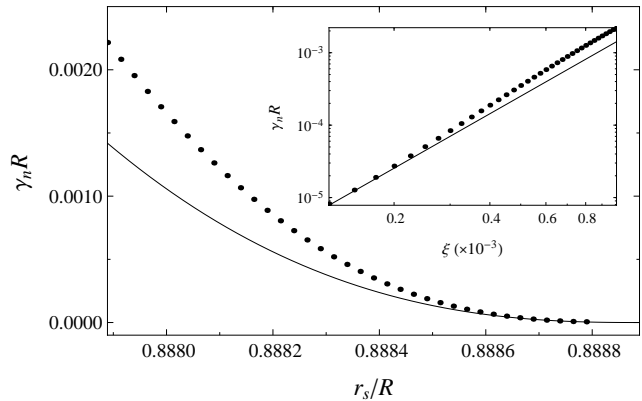}
\caption{Width of the $n=4$ resonance found numerically
by fitting the phaseshift $\delta $ (solid circles), and estimated from
Eqs.~\eref{eq:Energy} and \eref{eq:Widths1} (solid line), as a function of
$r_s/R$. The inset shows the same data as a function of $\xi$ on a double
logarithmic scale.
\label{fig:WidthPlot}}
\end{center}
\end{figure}

Thus we see that similar to the resonance energies $\eps _n$,
Eq.~(\ref{eq:Energy}), the resonance widths also decrease with $\xi $,
i.e., as the metric singularity is appoached. Taking Eq.~(\ref{eq:Energy})
into account we see from Eq.~(\ref{eq:Widths}) that for a fixed resonance
number $n$, $\gamma _n\propto \xi ^{5/2}$. This dependence is shown on
the inset of Fig.~\ref{fig:WidthPlot}.

The above derivation of the width underestimates the width (through the value
of $B$) by a factor $r_2^2/R^2\sim 0.1$, since the correct wavefunction
does not decrease through region III as quickly as that given by
Eq.~\eref{eq:IntNearBoundary} (see Fig.~\ref{fig:Wavefunction}). Restoring
this factor and evaluating $B$ numerically gives
\begin{equation}\label{eq:Widths1}
\gamma_n \simeq 28 \sqrt{\xi }\, \eps_{n}^{4}R^3,
\end{equation}
in good agreement with the numerical calculation, as shown by
Fig.~\ref{fig:WidthPlot}.


\section{Absorption Cross section}
In the limit $r_s\rightarrow 8R/9$ the short-range phaseshift and
the energy density of the resonances tend to infinity. Hence, neither
$\delta $, nor the resonance energies and widths, $\eps_n$ are $\gamma _n$
(which tends to zero) retain much physical meaning. However, as in the
near-black-hole metric case \cite{Resonances}, the cross section for capture
into these resonances, i.e., the effective absorption cross section, is well
defined.

Equation~(\ref{eq:Energy}) shows the level spacing between the resonances
is given by
\begin{equation}\label{eq:Spacing}
D \equiv \eps_{n+1} - \eps_n =\sqrt{2\xi }/R.
\end{equation}
Assuming a finite energy resolution $\Delta \eps \gg D$ and given that
$\gamma_n \ll D$, one can introduce the cross section for capture of the
particle into these long-lived states. According to the optical-model
considerations \cite{LLV3}, this {\em absorption} cross section is given by
\begin{equation}\label{eq:OpticalSigma}
\bar{\sigma}_{a}^{({\rm opt})}=\frac{2\pi^2}{\eps ^2}\,\frac{\gamma_n}{D }.
\end{equation}
Substituting the expression for $\gamma_n$, Eq.~(\ref{eq:Widths1}), into the
above expression yields
\begin{equation}\label{eq:siga}
\bar{\sigma}_{a}^{({\rm opt})}\simeq 390\eps^2 R^4.
\end{equation}
Unlike the low-energy absorption cross section for the black holes,
$\sigma _a=4\pi r_s^2$, the cross section (\ref{eq:siga}) vanishes at
zero energy. However, for non-zero incident energies the capture
capture cross section for massless particles is finite.

\section{Conclusions}
We have considered the problem of scattering of massless scalar particles
from a spherically symmetric constant density fluid sphere described by the Schwarzschild interior metric. 
We find that
despite the black hole limit being unattainable for such metric, a dense
spectrum of narrow resonances emerges in the limit $r_s\rightarrow 8R/9$
in which a singularity develops in the metric at the origin. This phenomenon
gives rise to a nonzero capture cross section of massless particles for
nonzero energis. This implies that an object that is not a black hole may
\textit{gravitationally} absorb particles making it appear black hole-like,
albeit with a different capture cross section.

\end{document}